\documentclass[10pt,letterpaper]{article}
\usepackage[top=0.85in,left=2.75in,footskip=0.75in]{geometry}

\usepackage{amsmath,amssymb}

\usepackage{changepage}

\usepackage[utf8x]{inputenc}

\usepackage{textcomp,marvosym}

\usepackage{cite}

\usepackage{nameref,hyperref}

\usepackage[right]{lineno}

\usepackage{microtype}
\DisableLigatures[f]{encoding = *, family = * }

\usepackage[table]{xcolor}

\usepackage{array}

\newcolumntype{+}{!{\vrule width 2pt}}

\newlength\savedwidth

\raggedright
\usepackage[aboveskip=1pt,labelfont=bf,labelsep=period,justification=raggedright,singlelinecheck=off]{caption}

\makeatletter
\renewcommand{\@biblabel}[1]{\quad#1.}
\makeatother

\usepackage{lastpage,fancyhdr,graphicx}
\usepackage{epstopdf}
\fancyheadoffset[L]{2.25in}
\fancyfootoffset[L]{2.25in}
\begin{document}
\vspace*{0.2in}

\begin{flushleft}
{\Large
\textbf\newline{Peer groups for organisational learning: clustering with practical constraints} 
}
\newline
\\
Daniel W. Kennedy\textsuperscript{1,2*},
Jessica Cameron\textsuperscript{1,2},
Paul P.-Y. Wu\textsuperscript{1,2},
Kerrie Mengersen\textsuperscript{1,2}
\\
\bigskip
\textbf{1} Science and Engineering Faculty, Queensland University of Technology, Brisbane, Queensland, Australia
\\
\textbf{2} ARC Centre for Excellence in Mathematical and Statistical Frontiers, Brisbane, Queensland, Australia
\\
\bigskip

%
%





* dan.w.kennedy@qut.edu.au

\end{flushleft}
\section*{Abstract}
Peer-grouping is used in many sectors for organisational learning, policy implementation, and benchmarking. 
Clustering provides a statistical, data-driven method for constructing meaningful peer groups, but peer groups must be compatible with business constraints such as size and
stability considerations.
Additionally, statistical peer groups are constructed from many different variables, and can be difficult to understand, especially for non-statistical audiences.
We developed methodology to apply business constraints to clustering solutions and allow the decision-maker to choose the balance between statistical goodness-of-fit and conformity to business constraints. 
Several tools were utilised to identify complex distinguishing features in peer groups, and a number of visualisations are developed to explain high-dimensional clusters for non-statistical audiences. 
In a case study where peer group size was required to be small ($\leq 100$ members), we applied constrained clustering to a noisy high-dimensional data-set over two subsequent years, ensuring that the clusters were sufficiently stable between years.
Our approach not only satisfied clustering constraints on the test data, but maintained an almost monotonic negative relationship between goodness-of-fit and stability between subsequent years.
We demonstrated in the context of the case study how distinguishing features between clusters can be communicated clearly to different stakeholders with substantial and limited statistical knowledge.



\section*{Introduction}
Organisational benchmarking has become increasingly widespread across industries, including health, education and government~\cite{Klassen2009}. 
An inherent consideration in any benchmarking exercise, which is suggested as a means for organisational learning and quality assurance, is that of comparing and contextualising an organisation's performance against its peers, known as peer-grouping~\cite{Segev1999,Hanning2007}. 
However, identifying relevant peer organisations is challenging due to the variability and uncertainty in the mix of clients and services provided by each organisation~\cite{Dimick2010,Wilson2008,Pope2009}. 
Although unsupervised clustering methods might appear as the solution to the peer-grouping problem, a key challenge to practical uptake lies in the management context where clustering outcomes need to be interpretable by stakeholders of varied backgrounds.
In addition, the goals of organisational learning and benchmarking implicitly require peer groups to remain sufficiently similar for long periods of time for such activities to occur.
Although the clustering literature is very well developed~\cite{Xu2009}, the unique needs of practical peer-grouping have yet to be specifically addressed.

\subsection*{Previous Work}

Peer groups have a strong similarity with the concept of strategic groups from industry organisation economics, which are defined as groups of companies with similar business models. They are operationally defined using variables relevant to industry, as groups with more in common than firms in different groups~\cite{Porter1980}.
Study of strategic groups has a long history dating back to the 1970's, with clustering being among the original techniques for identifying strategic groups~\cite{Cool1987}.
Clustering is characterised as the empirical method of identifying strategic groups against the theoretical methods which are expert-elicited, based on known qualitative differentiations between company strategies.

\subsubsection*{Inter-organisational learning}

One of the key goals of peer-grouping is peer mentoring, often referred to as inter-organisational learning, for two-way transmission of ideas and knowledge to improve performance~\cite[p.54]{Atterton2009}. 
A study has shown that mentoring activities from external peers are viewed favourably in the local government space~\cite{Woods2016} and it has been argued that constructive feedback provided by peers on strengths and weaknesses of an organisation is seen as more independent and encouraging to organisations~\cite{Wood2015}.
Although diversity in peer groups is recognised as being important for satisfactory peer mentoring, sufficient similarity is also required for real dialogue and trust~\cite[p. 105]{VANEWIJK2012}.
It has also been argued that to best facilitate the sharing of information and practices, peer groups must be stable (i.e. remain the same or similar) over time \cite{Bridges2000} and need to be a manageable size.

\subsubsection*{Clustering}

Clustering has been used to identify peer groups in contexts such as hospitals~\cite{Stefos1992} and educational institutions~\cite{Hurley2002}.
Clustering generally uses a data-derived measure of dissimilarity between pairs of observations to find groups where their pairwise dissimilarities are smaller within the group than for pairs of observations in different groups~\cite[p.502]{Hastie2009}.
For example, \cite{ZafraGmez2009}  employ k-means clustering to identify groups of similarly sized local government municipalities and socioeconomic demographics.
In contrast, hierarchical clustering is an example of an agglomerative approach, as organisations begin in separate clusters and are incrementally grouped into progressively larger clusters~\cite{Xu2009}.
Hierarchical clustering has also been employed for peer-grouping~\cite{Duan2014,Gmez2016} and suggested for its flexibility and simple structure.
In this approach, there is the potential for cluster size to be directly controlled; however, interpretability of clusters is not as straightforward as with a partitioning approach.

In general, clustering relies on discriminating characteristics such that organisations within a cluster are more similar and organisations in different clusters are more dissimilar (i.e. maximising discrimination between groups).
Organisational benchmarking contexts tend to involve many variables, creating a high-dimensional clustering problem~\cite{Klassen2009}.
As a result, cluster interpretability can be challenging, not just due to the curse of dimensionality, but also to potential non-linearities and statistical dependence/correlation between variables.
Visualisation and communication is also more challenging, although techniques such as the co-plot~\cite{Segev1999} have been used to explore organisational clusters and their relationship to different variables.

Of critical importance for clustering is determining a quantitative measure of pairwise dissimilarity between observations~\cite[p.502]{Hastie2009}.
Most often this dissimilarity is calculated based on differences for relevant variables, and then combined into a single values for each pair of observations.
Mixture models are probability models describing data which are derived from latent subgroups with different underlying distributions~\cite{Fruhwirth2006}. 
Gaussian mixture models have multivariate Gaussian distributions with subgroup-specific means and covariances between variables.
A popular non-parametric approach is the Dirichlet Process Mixture Model~\cite{Sethuraman1994}, which allows for an unknown number of subgroups, removing the need to pre-specify this number or fit multiple models with different numbers of subgroups.
Using Markov Chain Monte Carlo (MCMC) sampling, the probability, given the data that a pair of observations is in the same subgroup, can be estimated for all pairs and then used as the dissimilarity measure~\cite{Liverani2015}.

\subsubsection*{Practical Peer-grouping}

A common practical consideration in establishing peer groups to facilitate organisational learning is constraining the number of organisations in each group. Minimum size constraints have been previously explored for K-means clustering~\cite{Bradley2000}.
Most research, however, has focused on clustering with must-link and cannot-link constraints (for a recent example see \cite{Okabe2018}).
In addition, peer groups need to remain the same or similar over time~\cite{Bridges2000} to facilitate inter-organisational learning. Some recent work has gone into monitoring cluster stability~\cite{Namitha2019}, and earlier work has sought to characterise specific transitional behaviours~\cite{Ntoutsi2009,Oliveira2010}
However, in the context of peer-grouping, the need is to control stability rather than observe or analyse it.
Currently, peer-grouping that simultaneously addresses the need to produce clusters that reflect dissimilarities between organisations in the data, while maintaining consistency of clusters over time and achieving cluster size constraints, is an understudied problem.

In this paper, we develop a three-step methodology for aligning the statistical problem of clustering observations with the practical problem of peer-grouping, in the context of multivariate data.
In the first step, data is pre-processed and then used to inform a Dirichlet Process Mixture Model (DPMM), from which the main outputs are probabilistic estimates of similarity between observations.
In the second step, we use these similarity estimates and context-informed constraints to construct peer groups, which are both informed by data and usable in the application.
Finally, we present methods for identifying key variables and features which distinguish between the peer groups, and visualisations to communicate these distinctions to both technical and non-technical audiences.
Peer groups constructed through this methodology are justified statistically as having optimal goodness-of-fit while also being fit-for-purpose and understandable, which promotes trust in stakeholders.

\section*{Materials and methods}

In this section, the three-step methodology is outlined, and then described in detail (see Figure~\ref{fig:peer-grouping-3-step-diag}).
Since the main contributions of the work are the second step (construct peer groups) and third step (identify and communicate key features), more emphasis is placed on these steps.
With cluster analysis as the tool used to construct the peer groups from data, the terms `cluster' and `peer group' are used interchangeably throughout the manuscript.

\begin{figure}[!h]
	\caption{{\bf Diagram describing the 3-step methodology.}
		The business context, which are the features of the peer-grouping application, have an influence over clustering process, as well as the communication of the peer group features to analysts, policy-makers, and stakeholders.}
	\label{fig:peer-grouping-3-step-diag}
\end{figure}

The case study and data are described, along with initial summarising, transforming and rescaling of the data.
We use the well-known Dirichlet Process Mixture Model (DPMM) as the basis for computing a dissimilarity matrix as a representation of how similar organisations are to each other, based on the data.
We present three novel methods for constructing peer groups informed by both the data (via the dissimilarity matrix), and constraints on size and the presence of prior peer groups.
We use classification methods to automatically identify the important variables and groups of variables which define and discriminate between peer groups.
Finally, we present some standard and novel visualisations for communicating these clusters to different audiences, which in the context of peer groups includes different levels of statistical and numerical proficiency.

\subsection*{Case Study and Data}

The development of methodologies in this paper was motivated by a case study of a large organisation, with many outlets delivering a variety of services to clients.
The organisations to be peer-grouped here are the outlets.
The de-identified data comprise of many relevant factors for characterising differences between organisations and their client bases.
These include different measures of client demographics, such as age, socioeconomic status, and regionality (e.g. metropolitan, remote, or regional), as well as organisational characteristics, such as client numbers and frequency of services.
For reason of confidentiality, covariate names have been removed.
Two of the covariates are labelled `response 1' and `response 2'; however, these are not true responses and the clustering analysis is completely unsupervised.

Summary statistics such as the average client age, and demographic groups as represented by proportion of client base, were computed over the outlet's clients for the financial year.
Domain experts were consulted to reduce the number of variables down to 14 most important summary statistics.
These were expressed as continuous variables or proportions.
To improve comparability between variables, proportions were transformed onto the continuous scale using a logit transformation~\cite{Warton2011}, with 0 and 1 values shifted slightly towards the centre.
A small number of variables were highly right-skewed, so a log-transformation was applied.
Finally, all variables were scaled and centred post-transformation.

All analysis for this and subsequent steps was conducted in \textit{R}~\cite{R2020}. Full Rmarkdown analysis files are available online: \url{https://github.com/danwkenn/peer-group-clustering}.

\subsection*{Computing Dissimilarity using the DPMM}

Because the complexity of the data, a non-parametric Dirichlet Process Mixture Model (DPMM) method was fitted to cluster the data into peer groups for both years in which data were available: 2017 and 2018.
Separate DPMMs were fitted to the data for each year.
The output of this method was a posterior dissimilarity matrix (PDM), which gave the probability, under the DPMM, that a given pair of organisations were not in the same latent group.
The DPMM was chosen to compute a dissimilarity matrix over a standard distance method (e.g. Euclidean) because of the complexity of the data. The DPMM's flexibility allows for groups with different concentrations of observations, and is better able to deal with the high dimensionality of the covariate space, multicollinearity, non-linearity and non-Gaussian distributions, and actually assumes latent grouping in the data. An additional benefit of the DPMM is that both continuous and discrete variables can be used, although in this work we only explore continuous variables.

An initial analysis from this dissimilarity matrix used a standard clustering algorithm called Partitioning Around Medoids (PAM)~\cite{Kaufman1987} to identify the clusters, with the average silhouette width~\cite{Rousseeuw1987} used to achieve the optimal number of clusters.
These clusters, however, were too large based on the end-users' requirements for less than 100 organisations per peer group. There were substantial differences between the clusters for 2017 and 2018, which would result in large numbers of organisations being moved. In response, we developed two methods based on hierarchical clustering to create peer groups that satisfied this practical constraint, and a single method to reallocate organisations to new clusters while respecting the need for stability.

The DPMM was fitted using a Metropolis-within-Gibbs sampling algorithm as part of the \textit{PReMiuM} package~\cite{Liverani2015} in \textit{R}. The Markov Chain Monte Carlo (MCMC) sampler was run for 200,000 iterations on 6 parallel chains to fit DPMM.
This was done separately for both years to compute different PDMs. 
The goal of MCMC sampling is to approximate the Bayesian posterior distribution, with the specific goal in this context to accurately estimate elements of the PDM.
In order to assess whether convergence had been achieved, the MCMC chains were compared to assess whether (a) the chains had been run long enough to avoid Monte Carlo error, and (b) whether the chains had converged to the same stationary distribution.
The PDM estimates were compared between pairs of chains using scatter plots, and the distributions of PDM estimates were also compared using empirical density plots.
If convergence was achieved, then points in the scatter plots should fall on the line of equality and the densities would look very similar.
For the Dirichlet Process parameter $\alpha$, the trace, cumulative mean, autocorrelation, and density plots were used to assess convergence.
The (unnormalised) log-Posterior was also assessed for convergence using the same method as $\alpha$.

\subsection*{Clustering with Business Constraints}

The data-set in the case study presented the opportunity for the large organisation to construct data-driven peer groups for the purposes of organisational learning between organisations, developing defensible benchmarks, and tailoring policy to differentiated groups of organisations.
For these purposes, the peer groups were required to be at most 100 organisations in size and sufficiently stable over time.
The concern for stability arose because when organisations were reallocated to a new peer group for a new year, relationships developed for organisational learning would be compromised.
If policy is based on peer groups, frequent changes to peer groups cause discontinuity and uncertainty for stakeholders.
For benchmarking in particular, there is no size requirement, however, frequent changing of a stakeholder's peer group from year to year could reduce trust in any benchmark based on within-group comparison.

\subsubsection*{Maximum Cluster Size Constraint}

The first method, called \textit{kirigami-1}, uses the optimal clusters as an initial partition, then iteratively bisects partitions using the nodes of the hierarchical clustering tree, until all clusters have been cut into sub-clusters that are smaller than 100 entities.

The second method, called \textit{kirigami-2}, starts with all observations in singleton clusters and then proceeds in a similar manner as standard hierarchical clustering.
The difference is that if a merger results in a cluster which violates the upper threshold, this merger is blocked and the other possible mergers are considered instead. It proceeds until there are only two clusters or all possible mergers have been exhausted.

See \nameref{S1_Appendix} for the full algorithm pseudo-codes.

\subsubsection*{Temporal Stability Constraint}

To update the clusters in response to data for a new reporting period, a DPMM was again fitted to the new data to produce a corresponding dissimilarity matrix. 
We found that applying the same algorithm over successive years did not necessarily produce highly similar sets of peer groups.
Possible reasons for this are the due to noise in the data, although also potentially due to organisational `drift' in terms of client demographics.
These causes may be exacerbated by having clusters very close together.

To address this issue, we introduced a reallocation algorithm, which produces clusters with controllable levels of change in the peer groups between years.
The reallocation algorithm works similarly to kirigami-2 (see \nameref{S1_Appendix} for the pseudo-code).
It first identifies the top $p\times 100\%$ of observations with largest silhouette widths, reassigns them to singletons,  then proceeds with constrained hierarchical clustering as normal.
In reallocation, there are two competing objectives: goodness-of-fit and stability.
Goodness-of-fit is measured by clustering indices (see the next section for descriptions), while stability can be measured using cluster similarity indices.
However, the values of the similarity indices are difficult to interpret in context.

We considered that under a stable clustering, if two observations were in the same cluster (i.e. connected) in the previous year, then they should also be connected in the current year in most cases.
Therefore, a measure of stability is the proportion of connections in the previous year that are retained in the current year.
From this we developed a simple index called Proportion of Connections Retained (PCR):
\begin{equation}
	\text{PCR} = \frac{\text{No. of pairs connected in both years}}{\text{No. of pairs connected in the previous year}}
\end{equation}
The PCR is the proportion of connections present the previous year, which are retained in the current year. 
A connection is defined here between a pair of observations as being in the same cluster. 

The PCR is simpler to understand than the Rand Index~\cite{Rand1971} or Meila Index~\cite{Meila2007}, and in the context of peer-grouping, clarifies the cost of instability for a new set of peer groups, which may result in many connections between organisations being broken.
Furthermore, it implicitly includes the temporal order of the cluster solutions.
A negative aspect to the PCR is that it implicitly favours larger partitioning of entities into larger clusters, since merging clusters results in fewer lost connections.

To determine an optimal reclustering based on the current data (represented by the new dissimilarity matrix) and the previous peer groups, we adopted a grid-search approach.
The reallocation algorithm was run for a number of different evenly spaced values from $p=0$ to $p=0.95$.
Any potential clustering whose PCR was lower than the threshold was removed, and the clustering with the best goodness-of-fit index was selected.
A threshold of 90\% was chosen to represent a reasonable penalty, since this means that 1 in 10 connections would be broken each year by an observation being reallocated.

\subsubsection*{Clustering specifications}\label{sec:cluster-spec}

The kirigami methods are both based on hierarchical clustering, so we examined how group dissimilarity (as opposed to a single entity) should be calculated and how the optimal number of clusters should be determined.
We considered four methods for calculating group dissilimarity, known as linkage methods: average~\cite{Sokal1958}, Ward~\cite{Ward1963}, complete~\cite{Defays1977}, and single~\cite{Sibson1973} (see \nameref{S2_Appendix} for descriptions).
We were specifically interesed in whether the kirigami methods, combined with a particular linkage method, would produce better peer groups. 
These methods can all be efficiently calculated using the Lance-Williams algorithm, where during hierarchical clustering the dissimilarity for a merged cluster is a function of the dissimilarities being merged.

For calculating the optimal number of clusters, there are many measures in the literature~\cite{Vendramin2009}. We considered three measures which are often used: the average silhouette~\cite{Rousseeuw1987}, the Calinski-Haribaz (CH) Index~\cite{Calinski1974}, and the Pearson-Gamma measure~\cite{Halkidi2001}. These measures all indicate the optimal number of clusters with their maximum value.

The silhouette method measures for each observation the average distance between the observation's own cluster, and the minimum average distance to another cluster.
Hence, it measures the specificity of each observation to its own cluster. The average is calculated as the overall goodness-of-fit measure.
The CH index measures the ratio of the between-cluster and within-cluster sum-of-squares, so a good clustering by this measure has a large amount of variation explained by the clusters, similar to an Analysis-of-Variance problem.
Finally, the Pearson-Gamma measure calculates the Pearson correlation for a vector of all pairwise distances against a vector of ones and zeros, with 0 if the pair are in the same cluster or 1 if the pair are in different clusters.
This measures the linear correspondence between distance and the cluster equivalence relation. Full details and equations are described in \nameref{S2_Appendix}.

\subsection*{Identification of Key Variables and Features}

Since there were 14 input variables, clusters were located in a high-dimensional space.
It was therefore a complex task for an analyst to understand what specific variables and features were important for a given cluster in distinguishing it from other clusters.
We present tools based on well-known statistical methods to identify these features, and demonstrate their use in the case study data.
Variables identified in this step are then communicated in the next step for the simplified fingerprint plots.

Variables in the data are often correlated, resulting in the main components of variation being non-aligned with the main axes. These directions are well known as the main outputs of principal component analysis (PCA) and can be useful for identifying differences between clusters, since variation between clusters is likely to align along them. In addition to pairwise scatter-plotting of variables to examine cluster differences, this visualisation technique is used on the principal component scores of the data.

To accompany the visualisation approach described above, we also considered automatic methods of identifying variables which distinguish the clusters via classification algorithms.
To identify specific variables of importance, a random forest algorithm~\cite{Breiman2001} was run on the data with the cluster allocation as the response and the covariates as the predictors. This ranked the covariates in order of importance in distinguishing the clusters. We then trained the random forest algorithm on subsets of the data made up of two clusters to determine which variables were important in distinguishing between pairs of clusters. The benefit of the random forest importance is that because of the flexibility of the tree-based model, non-linear and interactive effects can be captured, as opposed to a logistic regression or other classification method. For a small number of trees ($<100$) the importance measure varied each time the random forest was fitted due to monte carlo error, so the number of trees was increased until the measure was sufficiently stable to reliably order the variables in order of importance.

Clusters can differ in terms of location or spread, and the differences can occur for a single variable, multiple variables or principal components. The random forest can identify an importance in the variable, however, we sought tools which could discriminate between these situations.

To identify whether location or spread were discriminative factors between two clusters, Linear Discriminant Analysis (LDA), Malanobis Distance (MD) and Quadratic Discriminant Analysis (QDA) were all fitted to the data and their predictive accuracy computed. LDA performs well when the locations of the clusters are distant compared to the spreads of the clusters, yet performs very poorly when the locations are similar as the model assumes equal spread for all clusters. The Malanobis Distance (MD) was computed to contrast with LDA, since using it for prediction assumes the same location for both clusters but different spreads. 
Finally, QDA allows for both differences in location and spread. Therefore, it should have the best prediction and forms an upper bound on accuracy for the other methods.

\subsection*{Communication to Stakeholders using Visualisation}

We sought to provide insights into both the data and the peer-grouping results.
The two stakeholder groups with whom we wanted to communicate were analysts and outlet representatives.
We assume analysts are familiar with standard visualisation tools such as boxplots and scatterplots.
In contrast, outlet representatives may prefer less statistical visualisations, and so graphics that summarised the information simply and succinctly were required.

Furthermore, the two groups likely have different questions about the data and the peer-grouping. Typical analyst question is: ``what makes a given peer group distinct from the other peer groups?'' To address this question, standard boxplots and scatterplots of the key variables and principal components were constructed to communicate distinctions between pair groups in one and two dimensions.

Typical outlet representative questions are more centred around the outlet and the relationship with its peer group:
\begin{enumerate}
	\item What is it about my outlet which makes peer group X right for me?
	\item What do the `typical' organisations look like for my peer group?
\end{enumerate}
To address the questions of this stakeholder group, visualisations of the key variables were constructed which required as little background knowledge as necessary in order to communicate distinctive features of peer groups.

Since the data are transformed prior to clustering, it may be difficult for a non-technical audience to understand the transformed scale.
Instead of plotting variables on the original scale or the transformed scale, the probability inverse transform (PIT) function~\cite{Angus1994} was used to transform each key variable in the data to be uniformly distributed between 0 and 1, essentially converting data-points into percentiles.
The overall distribution of the data in each variable was uniform, but for the subset of data in each cluster, this is not necessarily true.
Thus, the cluster distributions are  represented in terms of the cohort's data-distribution percentiles, which are readily understandable to the user. 

Further simplifying was done by splitting the data into 5 ordered bins which each contained 20\% of the observations.
We opted to use a 5-bin system as a good compromise between simplicity and representing the cluster's distribution, and a familiarity from its common usage in rating scales.
The resulting plot provides a means of understanding the distribution of peer groups compared to the cohort, without needing to know the underlying metric scale.

\section*{Results}

Initial analysis of the summary statistic data in 2017 found that a number of the variables were highly correlated with each other. These high correlations were also seen in the 2018 data.
Therefore, three variables which were highly correlated with other variables (based on the Variance Inflation Factor) were removed, resulting in 11 variables for clustering. 

\subsection*{Summary of Results}

Overall, kirigami-2 (bottom-up clustering) tended to produce better clusters according to goodness-of-fit metrics, compared to kirigami-1 (top-down clustering) for almost all upper-size thresholds (Section: \nameref{sec:kirig-12-comp}).
The CH index produced more reasonable cluster sizes and indicated clear preference on deciding the number of clusters.
There was not a clear distinction between different linkage methods, however, the single method tended to produce singleton clusters and the Ward method tended to produce larger clusters (Section:~\nameref{sec:linkage-comp}).
For the 2017 data, three clusters were found using the upper size threshold of 100.
Principal components and discrimination analyses showed a clear difference in centre between cluster 3 and the two other clusters, but the difference between clusters 1 and 2 was related to both centre and spread (Section:~\nameref{sec:hd-cluster-analysis}).
The fingerprint plot method was able to partially communicate these differences between clusters, but the differences between clusters 1 and 2 were not as clear (Section:~\nameref{sec:visualise}). 
The reallocation algorithm was able to find a compromise clustering solution between stability and goodness-of-fit when constructing new clusters for the 2018 data, using the 2017 peer groups as the original groupings, yet the relationship between stability and goodness-of-fit was not strictly monotonic (Section:~\nameref{sec:reallocation}).

\subsection*{Maximum cluster size constraint}\label{sec:kirig-12-comp}

Using the 2017 data, the behaviour of the constrained algorithms was observed for different upper-size thresholds with respect to cluster goodness-of-fit indices, cluster sizes and number (see Figure~\ref{fig:kirigami-index-v-thresh}).
For the three indices, there was a monotonic increase for both kirigami-1 and kirigami-2 as the size threshold was relaxed, with kirigami-2 being almost always better than kirigami-1 for all indices. 
Once the size threshold was above 160, the methods produced the exact same result, which was two clusters of size 168 and 32, with slight differences in these numbers depending on the linkage method used.
This behaviour was expected because these clusters correspond to the optimal clusters without constraint, so both methods are able to find this within the constrained solution space.

\begin{figure}[!h]
	\caption{{\bf Statistics for constrained cluster solutions as a function of upper-size threshold.}
		Index values (left column), cluster sizes and numbers (right column) for different values of the upper-size threshold for cluster size, for the two constrained clustering methods, kirigami-1 and kirigami-2. Cluster size and numbers are presented for solutions constructed using each of the three metrics, average silhouette width, CH, and Pearson-Gamma. Results here are presented for the average linkage method, however, the results did not vary markedly for other linkage methods.}
	\label{fig:kirigami-index-v-thresh}
\end{figure}

The greatest discrimination between the methods was seen for the CH index, where the kirigami-1 index was much lower for thresholds lower than 170 the index.
This was explained by the number of clusters, with kirigami-1 producing very large numbers of very small clusters.
Since every extra cluster was another `degree-of-freedom', the CH harshly penalised kirigami-1.
This contrasts with kirigami-2, which at most produced 10 clusters.
There was a much lower discrimination for the Pearson-Gamma index, with only a small advantage for kirigami-2 when the threshold was less than 100.

In terms of size, the largest cluster was always equal to the threshold, except once the optimal clusters could be found within the constrained space.
For kirigami-1, the smallest cluster size tended to be either 1, indicating one or more singleton clusters, or 32, corresponding to the optimal clusters, when the threshold was sufficiently high.
For kirigami-2, the cluster numbers and smallest sizes were dependent on the goodness-of-fit index used.
If the average silhouette width was used, then the size of the smallest cluster ranged from a singleton to 100, whereas if the Pearson-Gamma index was used, the smallest cluster only ranged between 1 and 6.
When the CH index was used, smallest cluster size was between 21 and 50.
This again shows the tendency for kirigami-2 and the CH index to prefer fewer, larger clusters, compared to the other indices and the kirigami-1 method, which promote many singleton or very small clusters.

\subsection*{Linkage Method and Number of Clusters}\label{sec:linkage-comp}

For kirigami-1, the number of clusters was chosen by the goodness-of-fit index, whereas the goodness-of-fit index was only used in kirigami-2 to decide the initial clustering, and successive cuts increased the number of clusters to satisfy the upper-size threshold constraint.
Therefore, it was of interest to investigate these indices' behaviours when changing the number of clusters for kirigami-1 specifically.
The upper-size threshold was fixed at 100 for the analysis in this section.

\begin{figure}
	\includegraphics[width=0.9\textwidth]{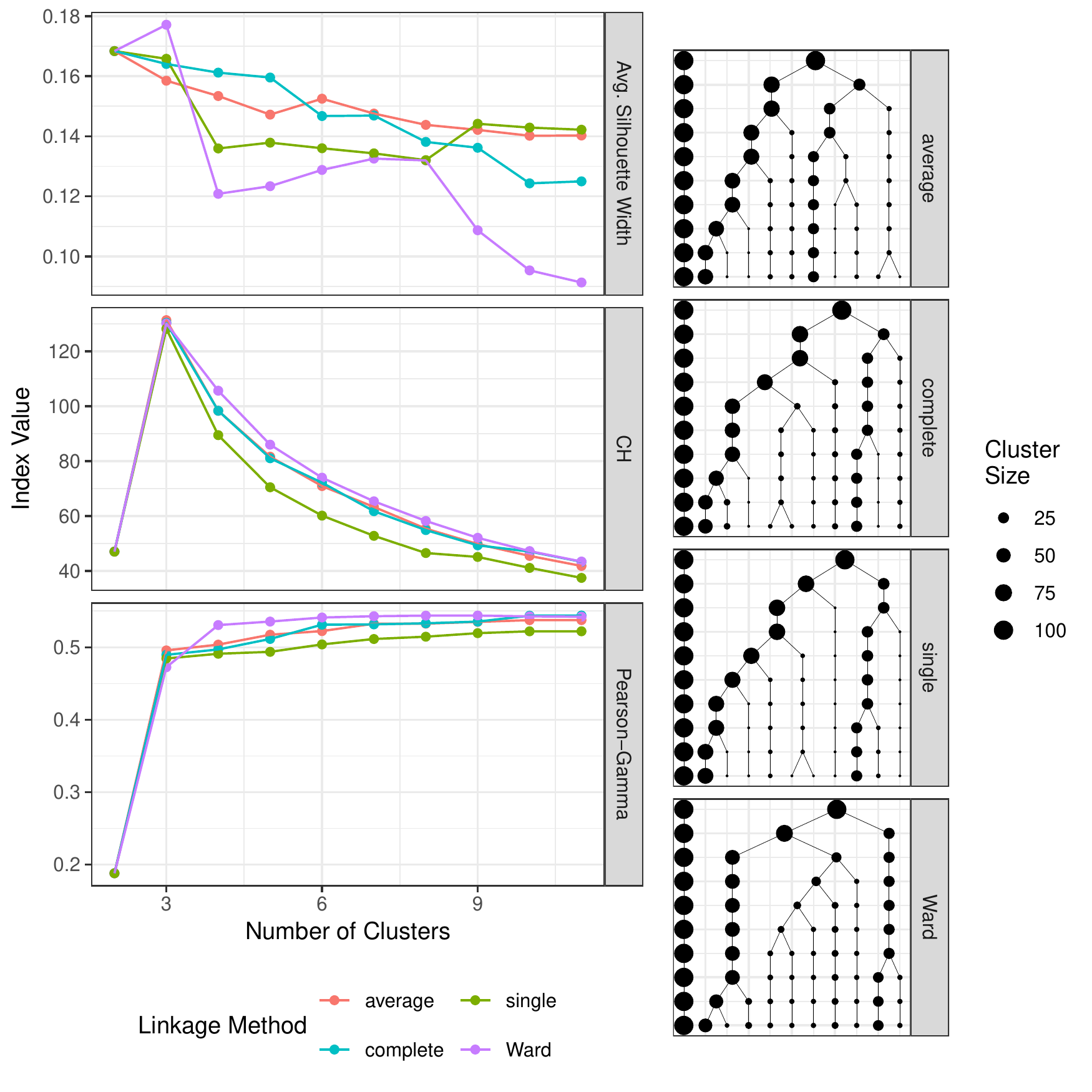}
	\caption{{\bf Goodness-of-fit and tree structures for different heirarchical clustering linkage methods.} (left) Clustering goodness-of-fit indices plotted against the number of clusters for kirigami-2. The algorithm was performed using four different types of linkage method. (right) The heirarchical trees constructed using the kirigami-2 algorithm for the four linkage methods. The trees are represented vertically, with the highest level representing 2 clusters, and the lowest level respresenting 11 clusters after 10 bisections. Each point represents a cluster, with the point size representing the size of the clusters. Clusters are connected to their `parent cluster' by vertical and diagonal edges.
	}
	\label{fig:kirigami-2-index-v-nclust}
\end{figure}

As shown in Figure~\ref{fig:kirigami-2-index-v-nclust}, the indices showed vastly different behavours.
The average silhouette width tended to reduce for larger numbers of clusters, with the highest value being 2 clusters for all linkage methods except for Ward, where the value was slightly higher for 3 clusters.
The Pearson-Gamma index instead showed an increase for all linkage methods as the number of clusters increased, plateauing after around 6 clusters.
The preference for 3 clusters was very clear in the CH index, with a sharp increase from 2 clusters to 3 and then a steady decline as more clusters were added.

The trees produced by the linkage methods were similar only for the first split and diverged quickly afterwards.
The Ward linkage method tended to result here in larger clusters, whereas in the other linkage methods it was more common for very small or singleton clusters to break off from a larger cluster.
Because the CH index penalises more for small clusters, the other linkage methods tended to produce poorer clusters according to this index.
By contrast, the average silhouette width index indicated the Ward clustering was the worst for four or more clusters.
The associated reduction from 0.17 to around 0.12 when the number of clusters was increased from 3 to 4 was actually the result of one cluster splitting into two new clusters, each of which was closer to the other clusters, resulting in a reduction in the silhouette widths of observations in the other two clusters.

\subsection*{Understanding High Dimensional Clusters}\label{sec:hd-cluster-analysis}

\begin{figure}
		\includegraphics[width=0.9\textwidth]{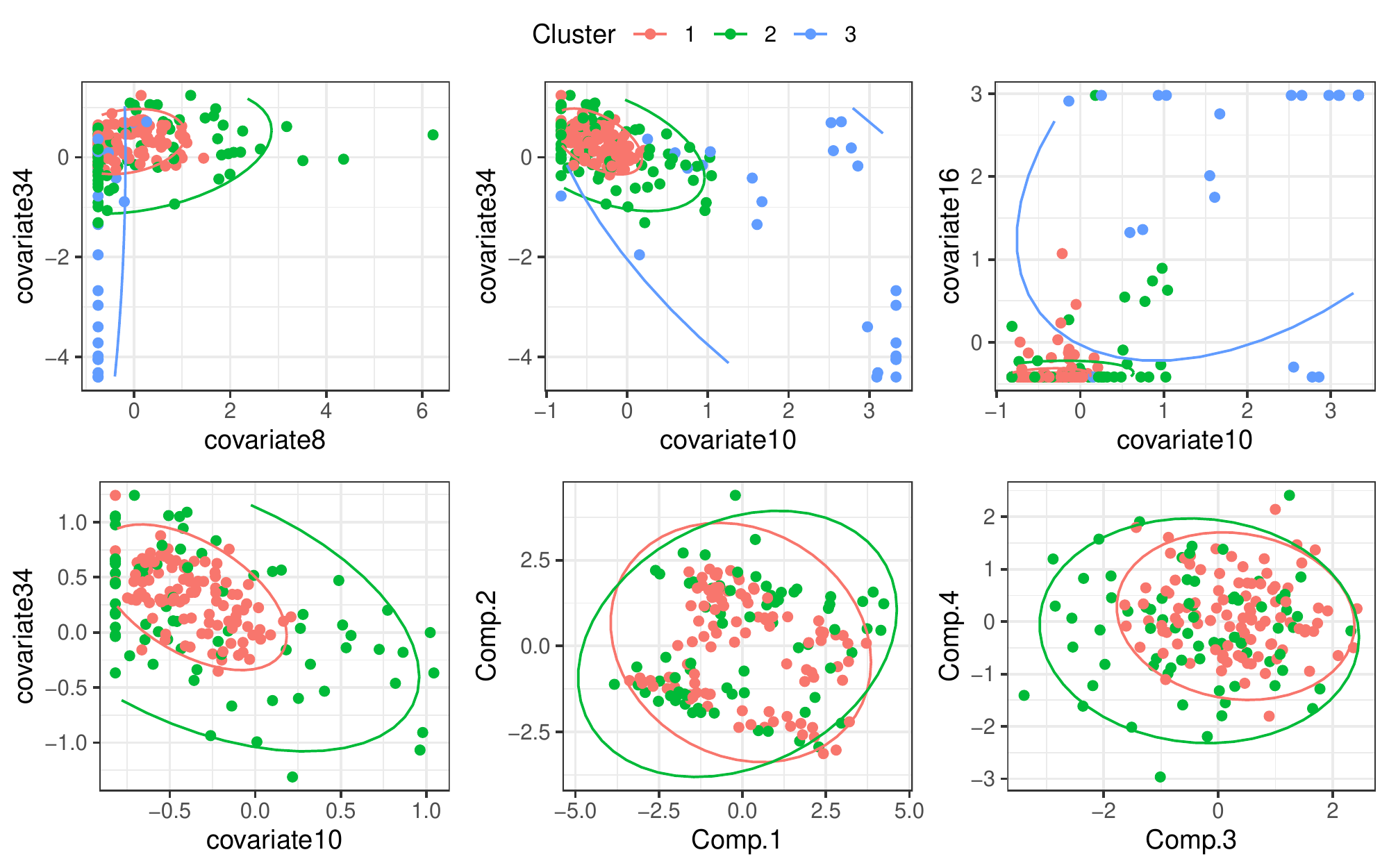}
	\caption{{\bf Pairwise scatter plots showing the difference between the clusterings for the 2017 data}. The top three scatter plots show all three clusters for three pairs of variables from the data, selected by random forest importance. The bottom three show only clusters 1 and 2, and show (bottom-left) the top two variables for discriminating between the two clusters and the first four principal component scores (bottom middle and left). Ellipses represent the 95\% probability regions, and are calculated based on the empirical means and covariance matrices of the clusters for a multivariate Gaussian distribution. The comparison of clusters 1 and 2 between covariate 10 and covariate 34 clearly shows the shell structure.
	}
	\label{fig:cluster-explore}
\end{figure}

It was found to be very easy to distinguish cluster 3 from clusters 1 and 2 in terms of a few variables.
Cluster 3's center was substantially separated from the other clusters, and it had relatively large spread (see Figure~\ref{fig:cluster-explore}).
It was much more difficult to identify differences between clusters 1 and 2, and standard plots were not useful for distinguishing between them.
Using random forest importance, the most important covariates for discriminating between them were covariate 34 and covariate 10, and plotting these revealed only small differences in the centre but a large difference in spread.
We termed this particular situation a `shell structure', since the observations in cluster 1 are surrounded in multi-dimensional space by the observations in cluster 2, which is much sparser.
From a PCA analysis we found that clusters 1 and 2 had heavy overlap in the main directions of variation in the data, although cluster 1 was more concentrated for some principal components.

Computing the Mahalanobis distance from the centre of the two clusters, it was found that cluster 2's observations tended to be further away than cluster 1's (83.9\% discrimination).
Linear Discriminant Analysis also produced a similar level of discrimination (84.5\%), indicating some difference between centre. 
Finally, a full quadratic discrimination analysis, which takes into account both centre and spread, was able to correctly classify between cluster 1 and 2 at 89\% accuracy. This shows there are small differences in both centre and spread.

\subsection*{Visualising Peer Groups}\label{sec:visualise}

\begin{figure}
	\includegraphics[width=0.9\textwidth]{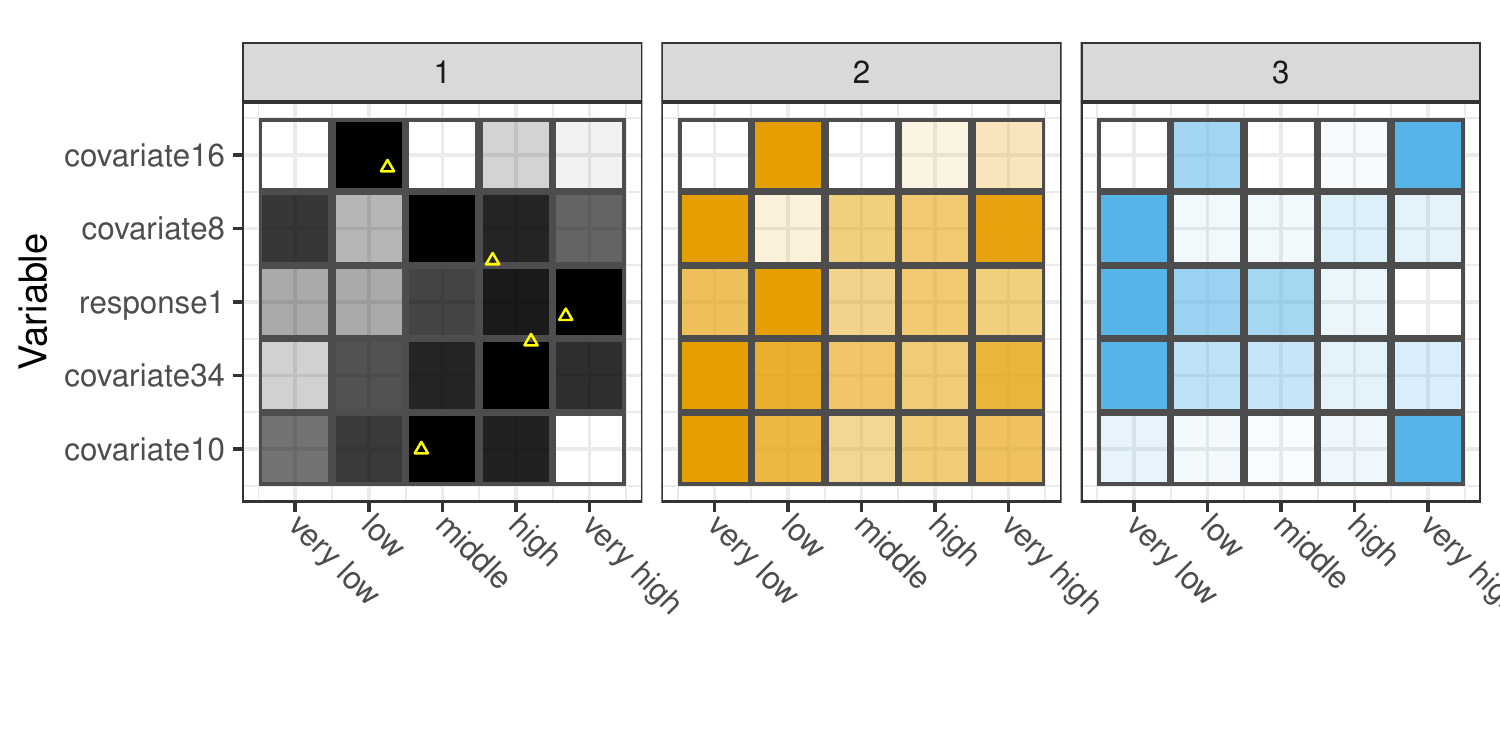}
	\caption{{\bf Fingerprint plots for the top 5 explanatory variables in the data-set, as chosen by the random forest discrimination}. Data are separated out into quintiles, and then apportioned out into clusters. The relative proportion of observations in each quintile is depicted by the transparency of the colour. Quintiles are given descriptors to guide the reader. A single observation in Cluster 1 is depicted with a yellow triangle to demonstrate how the corresponding stakeholder can interpret the graphic.}
	\label{fig:fingerprint-plots}
\end{figure}

The fingerprint plots (Figure~\ref{fig:fingerprint-plots}) show clear differences between the clusters, especially for comparing clusters 1 and 2 with cluster 3. Clusters 1 and 2 are more similar, as the figure suggests, however, there is a clear difference for covariate 10 and response 1.

Using the entire data distribution as a reference, the fingerprint plot can also show the individual cluster distributions. Although cluster 1 contains organisations in all quintiles for response 1, a greater proportion of its organisations tend to have high (highest 60-80\%) and very high (highest 80-100\%) response 1 values. Cluster 3 by contrast contains no large or very large organisations, but a large proportion of very small organisations (lowest 0-20\%).

The graphic also provides justification to the stakeholder that they have been correctly assigned to their cluster. The featured observation, chosen at random, is seen to be in typical bins for that cluster.
It is clear that it would be a poor fit for the other two clusters, particularly for covariate 34.

\subsection*{Reallocation}\label{sec:reallocation}

With the peer groups established for 2017, the posterior dissimilarity matrix for 2018 was used to reallocate observations to new peer groups when they were of sufficiently high silhouette distance from their 2017 peer group.
As with 2017, Ward's linkage method was used along with the CH index.

\begin{figure}
	\includegraphics[width=0.9\textwidth]{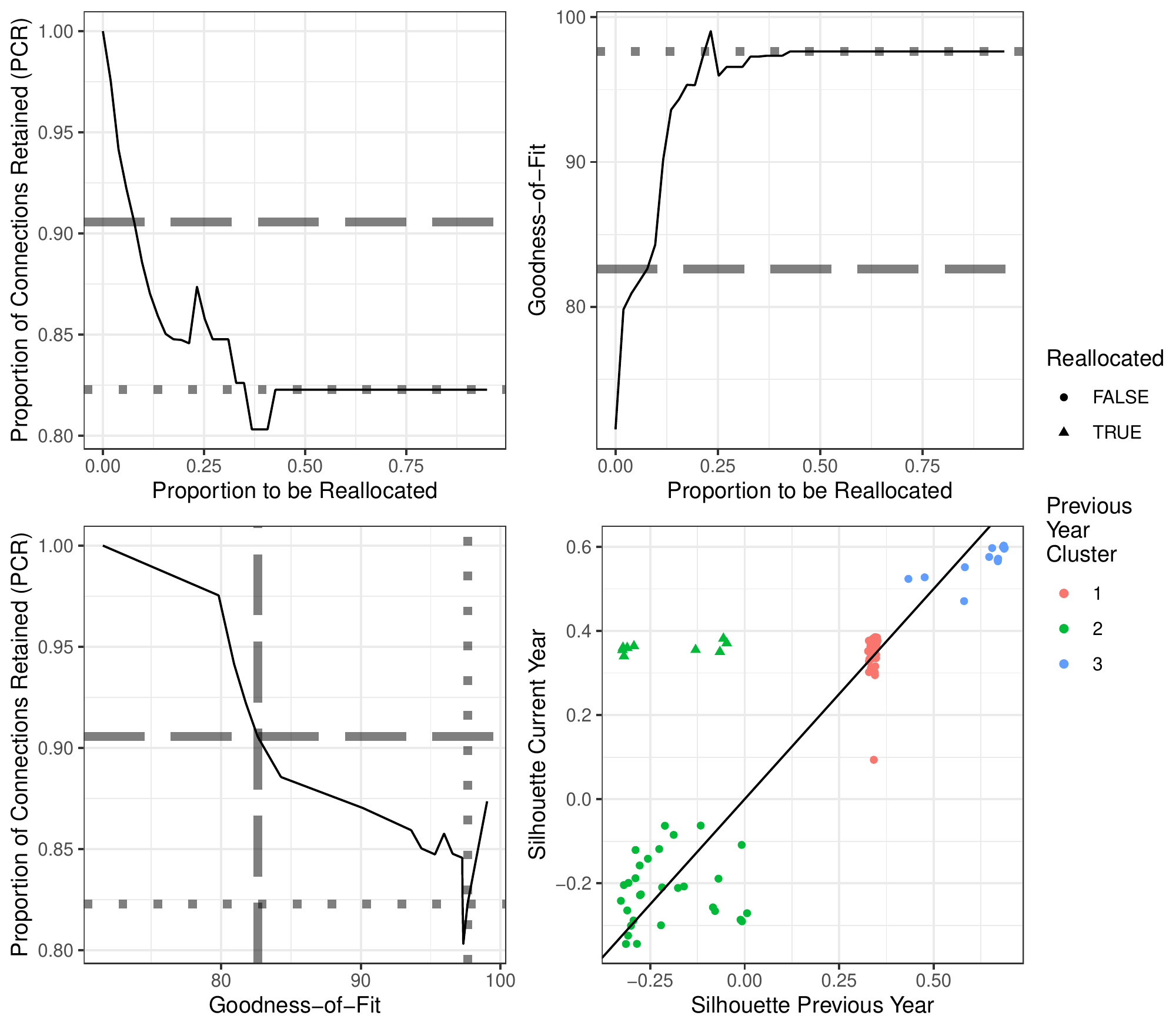}
	\caption{{\bf Goodness-of-fit and stability for different values of the reallocation threshold.} (top-left) The stability, as measured by the proportion of connections retained (PCR). (top-right) The CH index goodness-of-fit value for partitions created using different proportions of reallocated entities (bottom-left). The stability plotted against the goodness-of-fit, providing a line to evaluate the trade-off between the two objectives. In each of these three plots, the horizontal and vertical guidelines are provided to show the PCR and goodness of fit values for the reallocation clustering solution (dashed) based on a lower stability threshold of 90\% and a clustering solution with no stability constraint (dotted). (bottom-left) The silhouette distance in the previous year as a predictor of the silhouette distance in the following year}
	\label{fig:fit-versus-constraints}
\end{figure}

As expected, the stability of the cluster solutions measured by PCR reduced as more observations were assigned to be reallocated, and the goodness-of-fit, measured by the CH index, increased (see Figure~\ref{fig:fit-versus-constraints}). 
This increase was not monotonic in either case, as might be expected given increasing the reallocation threshold relaxed the stability constraint.
This was seen when the proportion to be reallocated was around 17\%, where the PCR briefly rose before continuing to decrease.
Investigation revealed this was because the number of clusters briefly reduced from 3 to 2, resulting in an increase in the PCR.
The number of clusters then changed again back to 3, with the original third cluster coalescing with a subset of observations from a larger cluster to form a new group.

The goodness-of-fit increased monotonically with the reallocation threshold with the exception of a single spike, which also coincided with the number of clusters changing from 3 to 2.
The increases plateaued quickly, as once the reallocation threshold was above 50\%, the optimal (size-constrained) solution could be reached and further relaxing of the stability constraint would not improve the fit.
There was a strong penalty for the stability threshold of 90\% based on the goodness-of-fit index. The CH-index for the reallocation cluster solution based on this penalty was 82.6, which was closer to the index for a clustering with no reallocation (71.5) than the solution with no constraint (97.6).

The reallocation solution for the $\text{PCR} < 90\%$, which was the solution chosen to be the basis for the 2018 peer groups, had 3 clusters that corresponded well to the 2017 clusters.
There were a total of 9 reallocated observations, all of which were from the 2017 peer group 2, and moved to both peer group 1 (7) and to peer group 3 (2). 
As shown in Figure~\ref{fig:fit-versus-constraints}, the silhouette distances were similar between years for observations that remained in the same cluster.
By contrast, the silhouette distances of the 9 reallocated observations, all from the 2017 peer group 2, increased substantially.
The silhouette distances for peer group 2 in both 2017 and 2018 were negative, a result of the shell structure that peer group 2 adopts around peer group 1.

\section*{Discussion}

In this paper, we developed methodology to assist the real-world peer-grouping problem through clustering.
Using a clustering approach instead of other methods such as expert elicitation has the clear benefit in being data-driven, thus reducing the risk of bias in deciding what separates organisations into peer groups.
Methods developed in this work allowed clustering solutions to be more amenable to the peer-grouping context by meeting the business constraints of size and stability.
The resulting clusters represented complex, multidimensional shapes, so classification algorithms were utilised to automatically identify important features and combinations of features, and identify differences between centre and spread.
Finally, a concise visualisation was developed to guide stakeholders in understanding peer group characteristics and  peer group similarities and differences.


The results show that the upper-size constraint forces a split to form clusters 1 and 2.
This causes a substantial loss of goodness-of-fit, and the reallocation method's constrained solution was also lower than the unconstrained goodness-of-fit.
While there may be improvements to be made on the kirigami and reallocation algorithms to find better constrained solutions, it does indicate that these are strong constraints which substantially reduce the distinction between clusters.
It is therefore important to consider from the perspective of the stakeholders how important the business constraints are, relative to the importance of having well-defined, data-driven peer groups.


The CH index showed a clear maximum at its optimal number of clusters, so was the most useful index to use in conjunction with the constrained clustering for these data.
The average silhouette was more sensitive to small differences in clustering solutions between the linkage methods.
Given these were only a small proportion of the data, this indicates that the average silhouette is dominated by a minority of observations.
The Pearson-Gamma method preferred many very small clusters, indicating it did not penalise for greater cluster number sufficiently.
Very small clusters which are dissimilar to all other larger clusters represent a challenge for the Pearson-Gamma index, which is the correlation between the dissimilarity and the cluster membership over all pairs of observations.
The merger of a very small cluster with the nearest cluster results in a reduction in the Pearson-Gamma index, but having very small clusters is undesirable for peer-grouping.
The CH index is therefore recommended here, but it is likely to be data-dependent as to which index provides the clearest preference for number of clusters. Context may dictate that a given index is more appropriate.

The kirigami-2 was overall a better performing algorithm for these data than the kirigami-1, regardless of the goodness-of-fit index used.
As shown in the results, the key reason for this was that the number of clusters produced by kirigami-1 was very high.
This is because kirigami-1 needs to unfold many cuts in the heirarchical clustering tree in order to reduce the size of clusters, which resulted in many singleton clusters being created in the process.
If the data were less noisy and clusters more clearly defined in the tree, then kirigami-1 would perform better, but this was not tested here.
On the other hand, the kirigami-2 algorithm produces its own tree, which sidesteps the need to use the unconstrained tree to find the clustering solution.
In practice therefore, kirigami-2 is likely to produce better defined peer groups.

The associated tree constructed by kirigami-2 contains numerous potential peer-grouping solutions with different numbers of clusters, all of which are subject to the required size constraint.
Providing multiple alternative solutions allows a practitioner greater flexibility in decision-making, given that peer-groups can often be subject to other requirements such as legal criteria, traditional business practices, or prior organisational relationships.
Since these requirements may not be easily coded into logical constraints on the solution, the kirigami-2 method may be embedded within a larger peer-grouping decision framework or extended to allow for soft constraints, more complex objective functions, or greater human input and judgement.

It was expected that in reallocation of observations, the goodness-of-fit would increase monotonically as the stability threshold was relaxed, but this was not strictly the case.
Instead there was some fluctuation when the stability threshold was close to being large enough to permit the optimal unconstrained clustering.
It is therefore recommended that this relationship is considered when using reallocation, rather than simply automatically running the algorithm with a chosen threshold value of the PCR.
This may be particularly important when the chosen threshold is small, meaning a greater proportion of observations are being reallocated.


Even with the tools demonstrated in this work, it remains difficult to explain complex, high dimensional clustering.
Domain knowledge may help to construct terms for the principal components of the data.
It may be possible to explain principal components in terms of abstract qualitative factors.
For example, component 4 in the case study had high loadings for number of clients and the number of sessions per client, in the same direction.
Therefore, component 4 might be described as `organisation size' to a stakeholder.
For example, the shell structure seen is very difficult to explain and justify as it defy basic intuition and 1D representation, as was seen for this case study.

The two peer-groups which make up the shell structure observed in the data were the result of the upper-size constraint, otherwise they would have been merged, but the constraint is not necessary for this behaviour to occur.
Comparing the performances of discriminant analysis algorithms and principal component analysis can dissect the differences between clusters in centre and spread, providing simple numerical summaries of the discrepancies between complex multidimensional shapes.
It nevertheless seems counter-intuitive when using single- and two-dimensional visualisations that two observations in the wider peer-group should be in the same peer-group, when they are both quite different to eachother and more similar to observations in the more dense peer-group.
The reasoning for the clustering solution is on the scale of the entire cohort rather than individual observations, because it is favourable to have a dense peer-group of close observations at the expense of a small number of observations in the sparser peer-group.
This stands as an example of where clustering results, particularly in high dimensions, are very difficult to explain in the peer-grouping context where clear, non-technical explanations are desired.

A peer-group with a large spread may mean organisations within are highly dissimilar to each other, but this is not necessarily a negative outcome with respect to benchmarking and inter-organisational learning.
For example, in the case of a shell structure, it could be reasonable to describe these organisations as atypical compared to the typical organisations in the denser, central peer-group.
Here, being atypical may either be a source of similarity and facilitate exchange between peer-groups, or indicate that the peer-group model may not be a valid method for these organisations to learn from one another.
Similarly, being atypical may also indicate that benchmarking is inappropriate for these organisations as they are not sufficiently similar to others.
Therefore, practitioners should evaluate by other means whether a wide peer-group can be used for a chosen purpose, but wideness does not preclude inter-organisational learning and may signal where benchmarking is inappropriate in the first place.


The critical challenges in merging the problem of peer-grouping with the problem of clustering are defining the business or contextual constraints on the clustering, and in the particular case of multi-dimensional data, developing a means for stakeholders to understand these constructed groups.
The first challenge is more suited to be solved by computational tools, but the constraints imposed on the clustering solution must be translated into thresholds or rules.
The second challenge is perhaps more difficult as it requires a complex, high-dimensional shape of a peer group to be reduced to single or two dimensions.
While it is tempting to instead construct peer groups with simplicity in mind, fully data-driven clustering helps to overcome analyst biases about which variables are most important for distinguishing organisations.
Instead, the second challenge presents the opportunity to develop novel visualisations, such as the fingerprint plot, which can bridge the gap between high-dimensional clustering and human understanding.


\section*{Acknowledgements}

Funding: KM and DK are supported by an ARC Laureate Fellowship (FL150100150).
The authors would like to acknowledge the agency which contributed financial support towards this research.

\section*{Supporting information}


\paragraph*{S1 Appendix.}
\label{S1_Appendix}
{\bf Pseudocode for constrained clustering algorithms.} Complete descriptions of the steps in the three algorithms presented (kirigami-1, kirigami-2, and reallocation).

\paragraph*{S2 Appendix.}
\label{S2_Appendix}
{\bf Descriptions of clustering specifications.} Descriptions of linkage criteria and cluster goodness-of-fit indices used in this research.

\nolinenumbers

%
%
%

\begin{thebibliography}{10}

\bibitem{Klassen2009}
Klassen A, Miller A, Anderson N, Shen J, Schiariti V,
  O{\textquotesingle}Donnell M.
\newblock Performance measurement and improvement frameworks in health,
  education and social services systems: a systematic review.
\newblock International Journal for Quality in Health Care. 2009;22(1):44--69.
\newblock doi:{10.1093/intqhc/mzp057}.

\bibitem{Segev1999}
Segev E, Raveh A, Farjoun M.
\newblock Conceptual maps of the leading {MBA} programs in the United States:
  core courses, concentration areas, and the ranking of the school.
\newblock Strategic Management Journal. 1999;20(6):549--565.
\newblock doi:{10.1002/(sici)1097-0266(199906)20:6<549::aid-smj39>3.0.co;2-f}.

\bibitem{Hanning2007}
Hanning BWT.
\newblock Length of stay benchmarking in the Australian private hospital
  sector.
\newblock Australian Health Review. 2007;31(1):150.
\newblock doi:{10.1071/ah070150}.

\bibitem{Dimick2010}
Dimick JB, Staiger DO, Birkmeyer JD.
\newblock Ranking Hospitals on Surgical Mortality: The Importance of
  Reliability Adjustment.
\newblock Health Services Research. 2010;45(6p1):1614--1629.
\newblock doi:{10.1111/j.1475-6773.2010.01158.x}.

\bibitem{Wilson2008}
Wilson D, Piebalga A.
\newblock Performance Measures, Ranking and Parental Choice: An Analysis of the
  English School League Tables.
\newblock International Public Management Journal. 2008;11(3):344--366.
\newblock doi:{10.1080/10967490802301336}.

\bibitem{Pope2009}
Pope DG.
\newblock Reacting to rankings: Evidence from
  {\textquotedblleft}America{\textquotesingle}s Best
  Hospitals{\textquotedblright}.
\newblock Journal of Health Economics. 2009;28(6):1154--1165.
\newblock doi:{10.1016/j.jhealeco.2009.08.006}.

\bibitem{Xu2009}
Xu R, Wunsch D.
\newblock Clustering.
\newblock Wiley-IEEE Press; 2009.

\bibitem{Porter1980}
Porter ME.
\newblock Competitive strategy : techniques for analyzing industries and
  competitors / Michael E. Porter.
\newblock Free Press New York; 1980.

\bibitem{Cool1987}
Cool KO, Schendel D.
\newblock Strategic Group Formation and Performance: The Case of the U.S.
  Pharmaceutical Industry, 1963{\textendash}1982.
\newblock Management Science. 1987;33(9):1102--1124.
\newblock doi:{10.1287/mnsc.33.9.1102}.

\bibitem{Atterton2009}
Atterton J, Thompson N, Carroll T.
\newblock Mentoring as a mechanism for improvement in local government.
\newblock Public Money {\&} Management. 2009;29(1):51--57.
\newblock doi:{10.1080/09540960802617368}.

\bibitem{Woods2016}
Woods R, Artist S, O'Connor G.
\newblock Learning in Australian local government: A roadmap for improving
  education {\&} training.
\newblock Commonwealth Journal of Local Governance. 2016; p. 108--126.
\newblock doi:{10.5130/cjlg.v0i18.4845}.

\bibitem{Wood2015}
Wood R, Tan S, Ryan R.
\newblock Councils learning from each other: An Australian case study,
  Australian Centre of Excellence for Local Government; 2015.

\bibitem{VANEWIJK2012}
van Ewijk E.
\newblock Mutual Learning in Dutch-Moroccan and Dutch-Turkish Municipal
  Partnerships.
\newblock Tijdschrift voor economische en sociale geografie.
  2012;103(1):101--109.
\newblock doi:{10.1111/j.1467-9663.2011.00698.x}.

\bibitem{Bridges2000}
Bridges J, Mazevska D, Pearse J.
\newblock Designing a nationally acceptable system of hospital peer grouping.
\newblock Australian Health Review. 2000;23(1):193.
\newblock doi:{10.1071/ah000193}.

\bibitem{Stefos1992}
Stefos T, LaVallee N, Holden F.
\newblock Fairness in prospective payment: a clustering approach.
\newblock Health services research. 1992;27(2):239—261.

\bibitem{Hurley2002}
Hurley RG.
\newblock Identification and Assessment of Community College Peer Institution
  Selection Systems.
\newblock Community College Review. 2002;29(4):1--27.
\newblock doi:{10.1177/009155210202900401}.

\bibitem{Hastie2009}
Hastie T, Tibshirani R, Friedman J.
\newblock The Elements of Statistical Learning.
\newblock Springer New York; 2009.
\newblock Available from: \url{https://doi.org/10.1007/978-0-387-84858-7}.

\bibitem{ZafraGmez2009}
Zafra-G{\'{o}}mez JL, L{\'{o}}pez-Hern{\'{a}}ndez AM, Hern{\'{a}}ndez-Bastida
  A.
\newblock Evaluating financial performance in local government: maximizing the
  benchmarking value.
\newblock International Review of Administrative Sciences. 2009;75(1):151--167.
\newblock doi:{10.1177/0020852308099510}.

\bibitem{Duan2014}
Duan X, ming Jin Z.
\newblock Positioning decisions within strategic groups.
\newblock Management Decision. 2014;52(10):1858--1887.
\newblock doi:{10.1108/md-08-2013-0415}.

\bibitem{Gmez2016}
G{\'{o}}mez J, Orcos R, Palomas S.
\newblock Do strategic groups explain differences in multimarket competition
  spillovers?
\newblock Strategic Organization. 2016;15(3):367--389.
\newblock doi:{10.1177/1476127016665358}.

\bibitem{Fruhwirth2006}
Fr{\"u}hwirth-Schnatter S.
\newblock Finite Mixture and Markov Switching Models.
\newblock Springer Series in Statistics. Springer New York; 2006.
\newblock Available from:
  \url{https://books.google.com.au/books?id=f8KiI7eRjYoC}.

\bibitem{Sethuraman1994}
Sethuraman J.
\newblock A CONSTRUCTIVE DEFINITION OF DIRICHLET PRIORS.
\newblock Statistica Sinica. 1994;4(2):639--650.

\bibitem{Liverani2015}
Liverani S, Hastie DI, Azizi L, Papathomas M, Richardson S.
\newblock {PReMiuM}: An {R} Package for Profile Regression Mixture Models Using
  Dirichlet Processes.
\newblock Journal of Statistical Software. 2015;64(7):1--30.

\bibitem{Bradley2000}
Bradley PS, Bennett KP, Demiriz A.
\newblock Constrained K-Means Clustering.
\newblock Redmond, WA: Microsoft Research; 2000.

\bibitem{Okabe2018}
Okabe M, Yamada S.
\newblock Clustering Using Boosted Constrained k-Means Algorithm.
\newblock Frontiers in Robotics and {AI}. 2018;5.
\newblock doi:{10.3389/frobt.2018.00018}.

\bibitem{Namitha2019}
Namitha K, Kumar GS.
\newblock A Framework for Monitoring Clustering Stability Over Time.
\newblock In: Emerging Research in Computing, Information, Communication and
  Applications. Springer Singapore; 2019. p. 467--477.
\newblock Available from: \url{https://doi.org/10.1007/978-981-13-5953-8_39}.

\bibitem{Ntoutsi2009}
Ntoutsi E, Spiliopoulou M, Theodoridis Y.
\newblock Tracing cluster transitions for different cluster types.
\newblock Control Cybern. 2009;38:239--259.

\bibitem{Oliveira2010}
Oliveira M, Gama Ja.
\newblock MEC --Monitoring Clusters' Transitions.
\newblock In: Proceedings of the 2010 Conference on STAIRS 2010: Proceedings of
  the Fifth Starting AI Researchers' Symposium. NLD: IOS Press; 2010. p.
  212–224.

\bibitem{Warton2011}
Warton DI, Hui FKC.
\newblock The arcsine is asinine: the analysis of proportions in ecology.
\newblock Ecology. 2011;92(1):3--10.
\newblock doi:{10.1890/10-0340.1}.

\bibitem{R2020}
{R Core Team}. R: A Language and Environment for Statistical Computing; 2020.
\newblock Available from: \url{https://www.R-project.org/}.

\bibitem{Kaufman1987}
Kaufman L, Rousseeuw PJ.
\newblock Clustering by means of Medoids.
\newblock In: Dodge Y, editor. Statistical Data Analysis Based on the
  $L_1$–Norm and Related Methods. The address of the publisher: Elsevier
  Science Pub. Co.; 1987. p. 405--416.

\bibitem{Rousseeuw1987}
Rousseeuw PJ.
\newblock Silhouettes: A graphical aid to the interpretation and validation of
  cluster analysis.
\newblock Journal of Computational and Applied Mathematics. 1987;20:53--65.
\newblock doi:{10.1016/0377-0427(87)90125-7}.

\bibitem{Rand1971}
Rand WM.
\newblock Objective Criteria for the Evaluation of Clustering Methods.
\newblock Journal of the American Statistical Association.
  1971;66(336):846--850.
\newblock doi:{10.1080/01621459.1971.10482356}.

\bibitem{Meila2007}
Meil{\u{a}} M.
\newblock Comparing clusterings{\textemdash}an information based distance.
\newblock Journal of Multivariate Analysis. 2007;98(5):873--895.
\newblock doi:{10.1016/j.jmva.2006.11.013}.

\bibitem{Sokal1958}
Sokal RR, Michener CD.
\newblock A statistical method for evaluating systematic relationships.
\newblock University of Kansas Science Bulletin. 1958;38:1409--1438.

\bibitem{Ward1963}
Ward JH.
\newblock Hierarchical Grouping to Optimize an Objective Function.
\newblock Journal of the American Statistical Association.
  1963;58(301):236--244.
\newblock doi:{10.1080/01621459.1963.10500845}.

\bibitem{Defays1977}
Defays D.
\newblock An efficient algorithm for a complete link method.
\newblock The Computer Journal. 1977;20(4):364--366.
\newblock doi:{10.1093/comjnl/20.4.364}.

\bibitem{Sibson1973}
Sibson R.
\newblock {SLINK}: An optimally efficient algorithm for the single-link cluster
  method.
\newblock The Computer Journal. 1973;16(1):30--34.
\newblock doi:{10.1093/comjnl/16.1.30}.

\bibitem{Vendramin2009}
Vendramin L, Campello RJGB, Hruschka ER.
\newblock On the Comparison of Relative Clustering Validity Criteria.
\newblock In: Proceedings of the 2009 {SIAM} International Conference on Data
  Mining. Society for Industrial and Applied Mathematics; 2009.Available from:
  \url{https://doi.org/10.1137/1.9781611972795.63}.

\bibitem{Calinski1974}
Calinski T, Harabasz J.
\newblock A dendrite method for cluster analysis.
\newblock Communications in Statistics - Theory and Methods. 1974;3(1):1--27.
\newblock doi:{10.1080/03610927408827101}.

\bibitem{Halkidi2001}
Halkidi M, Batistakis Y, Vazirgiannis M.
\newblock Journal of Intelligent Information Systems. 2001;17(2/3):107--145.
\newblock doi:{10.1023/a:1012801612483}.

\bibitem{Breiman2001}
Breiman L.
\newblock Machine Learning. 2001;45(1):5--32.
\newblock doi:{10.1023/a:1010933404324}.

\bibitem{Angus1994}
Angus JE.
\newblock The Probability Integral Transform and Related Results.
\newblock SIAM Review. 1994;36(4):652--654.

\end{thebibliography}

\end{document}


\maketitle
	
	\begin{algorithm}
		\begin{algorithmic}[1]
			\State Perform standard heirarchical clustering to get a series of partitions forming a tree: $\left\{\mathcal{P}^{(0)},\mathcal{P}^{(1)},...,\mathcal{P}^{(n)}\right\}$, with $\mathcal{P}^{(0)} = \left\{\left\{1,...,n\right\}\right\}$ and $\mathcal{P}^{(n)} = \left\{\left\{1\right\},\left\{2\right\},...,\left\{n\right\}\right\}$.
			\State For some goodness-of-fit index $G$,  $t_{\text{max}} := \text{argmax}_t{\left(G{\left(\mathcal{P}^{(t)}\right)}\right)}$
			\State $\mathcal{P} := \mathcal{P}^{\left(t_{\text{max}}\right)}$.
			\While{$\exists{P}\in \mathcal{P}$ s.t. $\left|P\right| > \lambda$}
			\State Find the smallest $t$ s.t. there exist a $P^{(t)}_p,P^{(t)}_q\in \mathcal{P}^{(t)}$ where $P = P^{(t)}_p \cup P^{(t)}_q$.
			\State $\mathcal{P} := \left(P/\left\{P\right\}\right) \cup \left\{P^{(t)}_p,P^{(t)}_q\right\}$.
			\EndWhile
			\State Return $\mathcal{P}$.
		\end{algorithmic}
		\caption{Kirigami-1: Top-down constrained clustering for an upper cluster size threshold of $\lambda$.}\label{alg:kirig-1}
	\end{algorithm}
	
	\begin{algorithm}
		\begin{algorithmic}[1]
			\State Initialise the partition $\mathcal{P}^{(0)}:= \left\{\left\{1\right\},...,\left\{n\right\}\right\}$, and dissimilarity matrix $D:=d{\left(\mathcal{P}^{(0)}\right)}$.
			\State $t := 0$.
			\While{$\exists{p,q}$ s.t. $p \neq q$, $D_{p,q} < \infty$, and $\left|\mathcal{P}^{(t)}\right| > 1$}
			\State $\left(p,q\right) := \text{argmin}_{p,q}{D_{p,q}}$.
			\If{$\left|P^{(t)}_p\cup P^{(t)}_q\right| > \lambda$}
			\State $D_{p,q} := \infty$.
			\Else
			\State $t:=t+1$
			\State $\mathcal{P}^{(t)} := \left(\mathcal{P}^{(t-1)}/\left\{P^{(t)}_p,P^{(t)}_q\right\}\right)\cup \left\{P^{(t)}_p\cup P^{(t)}_q\right\}$
			\State $D = d{\left(\mathcal{P}^{(t)}\right)}$
			\EndIf
			\EndWhile
			\State For some goodness-of-fit index $G$,  $t_{\text{max}} := \text{argmax}_t{\left(G{\left(\mathcal{P}^{(t)}\right)}\right)}$
			\State Return $\mathcal{P}^{\left(t_{\text{max}}\right)}$.
		\end{algorithmic}
		\caption{Kirigami-2: Bottom-up constrained clustering for an upper cluster size threshold of $\lambda$. Dissimilarity function $d$ is defined by the type of heirarchical clustering.}\label{alg:kirig-2}
	\end{algorithm}

	\begin{algorithm}
		\begin{algorithmic}[1]
			\State Allocate observations in $\mathcal{O}$ to based on previous year's partition $\mathcal{P}^{(0)} := \mathcal{P}^{\text{previous year}}$.
			\State Let $O_L$ be the set of $\lfloor p N \rfloor$ observations with the highest $\delta{\left(i,\mathcal{P}\right)}$ values.
			\State $\mathcal{P}^{(0)} := \left\{P/O_L : P \in \mathcal{P}^{(0)}\right\} \cup \left\{\left\{i\right\}:i\in O_L\right\} \cup \left\{\left\{i\right\}:i\in \mathcal{N}\right\}$.
			\State Compute dissimilarity matrix $D:=d{\left(\mathcal{P}^{(0)}\right)}$.
			\State $t := 0$.
			\While{$\exists{p,q}$ s.t. $p \neq j$, $D_{p,q} < \infty$, and $\left|\mathcal{P}^{(t)}\right| > 1$}
			\State $\left(p,q\right) := \text{argmin}_{p,q}{D_{p,q}}$.
			\If{$\left|P^{(t)}_p\cup P^{(t)}_q\right| > \lambda$}
			\State $D_{p,q} := \infty$.
			\Else
			\State $t:=t+1$
			\State $\mathcal{P}^{(t)} := \left(\mathcal{P}^{(t-1)}/\left\{P^{(t)}_p,P^{(t)}_q\right\}\right)\cup \left\{P^{(t)}_p\cup P^{(t)}_q\right\}$
			\State $D = d{\left(\mathcal{P}^{(t)}\right)}$
			\EndIf
			\EndWhile
			\State For some goodness-of-fit index $G$,  $t_{\text{max}} := \text{argmax}_t{\left(G{\left(\mathcal{P}^{(t)}\right)}\right)}$
			\State Return $\mathcal{P}^{\left(t_{\text{max}}\right)}$.
		\end{algorithmic}
		\caption{Reallocation: Hierarchical clustering based on a percentage $p\times 100\%$ of (potential) leavers and an upper cluster size threshold of $\lambda$. Number of observations is given by $N$. Set of old observations is given by $\mathcal{O}$, and set of new observations is given by $\mathcal{N}$. Dissimilarity function $d$ is defined by the type of heirarchical clustering. Distance metric for each observation $i$ to their assigned cluster is given by $\delta{\left(i,\mathcal{P}\right)}$.}\label{alg:reallocation}
	\end{algorithm}


\maketitle
	
	\begin{table}[h]
		\caption{Description of the four linkage criteria for heirarchical agglomerative clustering. The equation computes the dissimilarity between a cluster $C_3$ and the agglomeration of two other clusters $C_1$ and $C_2$. The dissimilarities between clusters are denoted by $D_{\cdot,\cdot}$ and the dissimilarities between single entities are denoted by $d_{\cdot,\cdot}$. Cluster sizes are given by $n_1$, $n_2$ and $n_3$.}\label{tab:linkage-criteria}
		{
			\begin{tabular}{lll}
			\hline
			Linkage Type & Description & Equation\\\hline
			\rule{0pt}{5ex}single & \parbox{0.5\linewidth}{Dissimilarity between two clusters is the minimum pairwise dissimilarity between one element from each cluster.} & $\text{min}{\left(\left\{d_{i,j}:i\in C_1\cup{}C_2\right\}\right)}$ \\\hline
			\rule{0pt}{5ex}complete & \parbox{0.5\linewidth}{Dissimilarity between two clusters is the maximum pairwise dissimilarity between one element from each cluster.} & $\text{max}{\left(\left\{d_{i,j}:i\in C_1\cup{}C_2\right\}\right)}$ \\\hline
			average & \parbox{0.5\linewidth}{Dissimilarity between two clusters is the average pairwise dissimilarity between one element from each cluster, computed over all possible such pairs.} & $\frac{\sum_{i\in{C_1\cup{}C_2},j\in{C_3}}{d_{i,j}}}{\left(n_1 + n_2\right)n_3}$ \\\hline
			Ward & \parbox{0.5\linewidth}{When distance is Euclidean, dissimilarity is the increase in variance for the cluster being merged.} & $\begin{matrix}\frac{n_1 + n_3}{n_1 + n_2 + n_3}D_{1,3} + \frac{n_2 + n_3}{n_1 + n_2 + n_3}D_{2,3} \\-\frac{n_3}{n_1 + n_2 + n_3}D_{1,2}\end{matrix}$ \\\hline
		\end{tabular}
	}
	\end{table}

	\begin{table}
		\caption{Descriptions of the three clustering criteria considered in this research. The numbers of observations and clusters are $N$ and $K$ respectively. The arithmetic mean of a given cluster $k$ is given by $\bar{\mathbf{x}}_k$, and the overall mean is given by $\bar{\mathbf{x}}$.}\label{tab:clustering-criteria}
		\begin{tabular}{p{2cm}ll}
			\hline
			Name & Description & Equation \\\hline
			Average Silhouette Width & \parbox{0.5\linewidth}{The silhouette width is a measure defined for each observation $i$ which compares the average distance between $i$ and all other observations in the same cluster, $a{(i)}$, with the average distance between $i$ and the closest other cluster, $b{(i)}$.\\The silhouette width is calculated as \[s{(i)} = \left(b{(i)} - a{(i)}\right)/ \text{max}{\left(b{(i)} , a{(i)}\right) }.\]  The average silhouette width is the arithmetic mean over the entire data-set. }& $\frac{1}{N}{\sum_{i = 1}^{N}{s{(i)}}}$\\\hline
			\parbox{0.15\linewidth}{Calinski-Haribasz (CH) Index} & \parbox{0.5\linewidth}{Calculated as the ratio of the within-cluster variance and the between cluster variance, with a correction for ``degrees of freedom''. Highly similar to the $F$-statistic in Analysis-of-Variance.} & $\frac{N-k}{k-1}\frac{\sum_{k = 1}^{K}{\sum_{i,j\in C_k, i\neq j}{d{\left(\mathbf{x}_i,\bar{\mathbf{x}}_k\right)}^2}}}{\sum_{k=1}^{K}{d{\left(\bar{\mathbf{x}}_k,\bar{\mathbf{x}}\right)}^2}}$\\\hline
			\parbox{0.15\linewidth}{Pearson-Gamma} & \parbox{0.5\linewidth}{Equal to the Pearson correlation over each pair of observations $i$,$j$ between the pairwise distance $d_{ij}$ and the indicator $\text{I}{\left(i \sim j\right)}$.} & As described.\\\hline
		\end{tabular}
	\end{table}